\documentclass[showpacs,floatfix,citeautoscript,twocolumn,prb,superscriptaddress]{revtex4-1}

\usepackage{graphicx,xcolor,upgreek}
\usepackage{amsmath}
\usepackage{comment}
\usepackage{textcomp}
\usepackage{inputenc}

\newcommand{\kk}[1]{\textcolor{green}{{\bf Kisung: #1 }}}

\begin{document}

\title{Strongly two-dimensional exchange interactions in the in-plane metallic antiferromagnet Fe$_2$As probed by inelastic neutron scattering}

\author{Manohar H.\ Karigerasi}
\author{Kisung Kang}
\affiliation{Department of Materials Science and Engineering and Materials Research Laboratory, University of Illinois at Urbana-Champaign, Urbana, IL 61801, USA}
\author{Garrett E. Granroth}
\affiliation{Neutron Scattering Division, Oak Ridge National Laboratory, Oak Ridge, TN 37831, USA}
\author{Arnab Banerjee}
\affiliation{Neutron Scattering Division, Oak Ridge National Laboratory, Oak Ridge, TN 37831, USA}
\affiliation{Department of Physics and Astronomy, Purdue University, West Lafayette, IN 47907, USA}
\author{Andr\'{e} Schleife}
\affiliation{Department of Materials Science and Engineering and Materials Research Laboratory, University of Illinois at Urbana-Champaign, Urbana, IL 61801, USA}
\affiliation{National Center for Supercomputing Applications, University of Illinois at Urbana-Champaign, Urbana, IL 61801, USA}
\author{Daniel P. Shoemaker}\email{dpshoema@illinois.edu}
\affiliation{Department of Materials Science and Engineering and Materials Research Laboratory, University of Illinois at Urbana-Champaign, Urbana, IL 61801, USA}


\begin{abstract}

To understand spin interactions in materials of the Cu$_2$Sb structure type, inelastic neutron scattering of Fe$_2$As single crystals was examined at different temperatures and incident neutron energies. The experimental phonon spectra match well with the simulated phonon spectra obtained from density functional theory (DFT) calculations. The measured magnon spectra were compared to the simulated magnon spectra obtained via linear spin wave theory with the exchange coupling constants calculated using the spin polarized, relativistic Korringa-Kohn-Rostoker method in Zhang \emph{et al}. (2013). The simulated magnon spectra broadly agree with the experimental data although, the energy values are underestimated along the $K$ direction. Exchange coupling constants between Fe atoms were refined by fits to the experimental magnon spectra, revealing stronger nearest neighbor Fe1-Fe1 exchange coupling than previously reported. The strength of this exchange coupling is almost an order of magnitude higher than other exchange interactions despite the three-dimensional nature of the phonon interactions. The lack of scattering intensity at energies above 60~meV makes unconstrained determination of the full set of exchange interactions difficult, which may be a fundamental challenge in metallic antiferromagnets.


\end{abstract}

\maketitle 

\section{Introduction}

With recent interest towards understanding the possibility of electrical switching behavior in metallic antiferromagnets,\cite{Baltz2018,Siddiqui2020,Jungfleisch2018,Zelezny2018} notably in CuMnAs\cite{Wadley2016,Grzybowski2017,Wadley2018,Matalla-Wagner2019} and Mn$_2$Au,\cite{Meinert2018,Bodnar2019} the relationships between their static magnetic orders,\cite{Wadley2013,Hills2015,Wadley2015,Saidl2017} in some cases are quite recently determined, and their spin dynamics \cite{Grzybowski2017,Bodnar2019,Yang2019,Yang2020} are of crucial interest. CuMnAs is a member of a larger family of easy-plane metallic antiferromagnets in the Cu$_2$Sb structure type,\cite{Nateprov2011,Wadley2013} which includes Cr$_2$As,\cite{Yuzuri1960} Mn$_2$As,\cite{Yuzuri1960Mn2As} and Fe$_2$As.\cite{Katsuraki1966}
The proposed switching involves a field-like torque from exchange interactions between the carrier spins and the moments of the magnetic atoms. The non-equilibrium current-induced spin polarization is staggered across the two sublattices and exerts a uniform torque on the N\'{e}el vector.\cite{Zelezny2014,Zelezny2017,Wadley2016}
While the static spin arrangements of these easy-plane antiferromagnets are known, the underlying energy scales and dynamics are less so. 
Determination of fundamental exchange and anisotropy energies are essential to understand what energy barriers and resonances may dominate in these materials.

\begin{figure}
\centering\includegraphics[width=0.85\columnwidth]{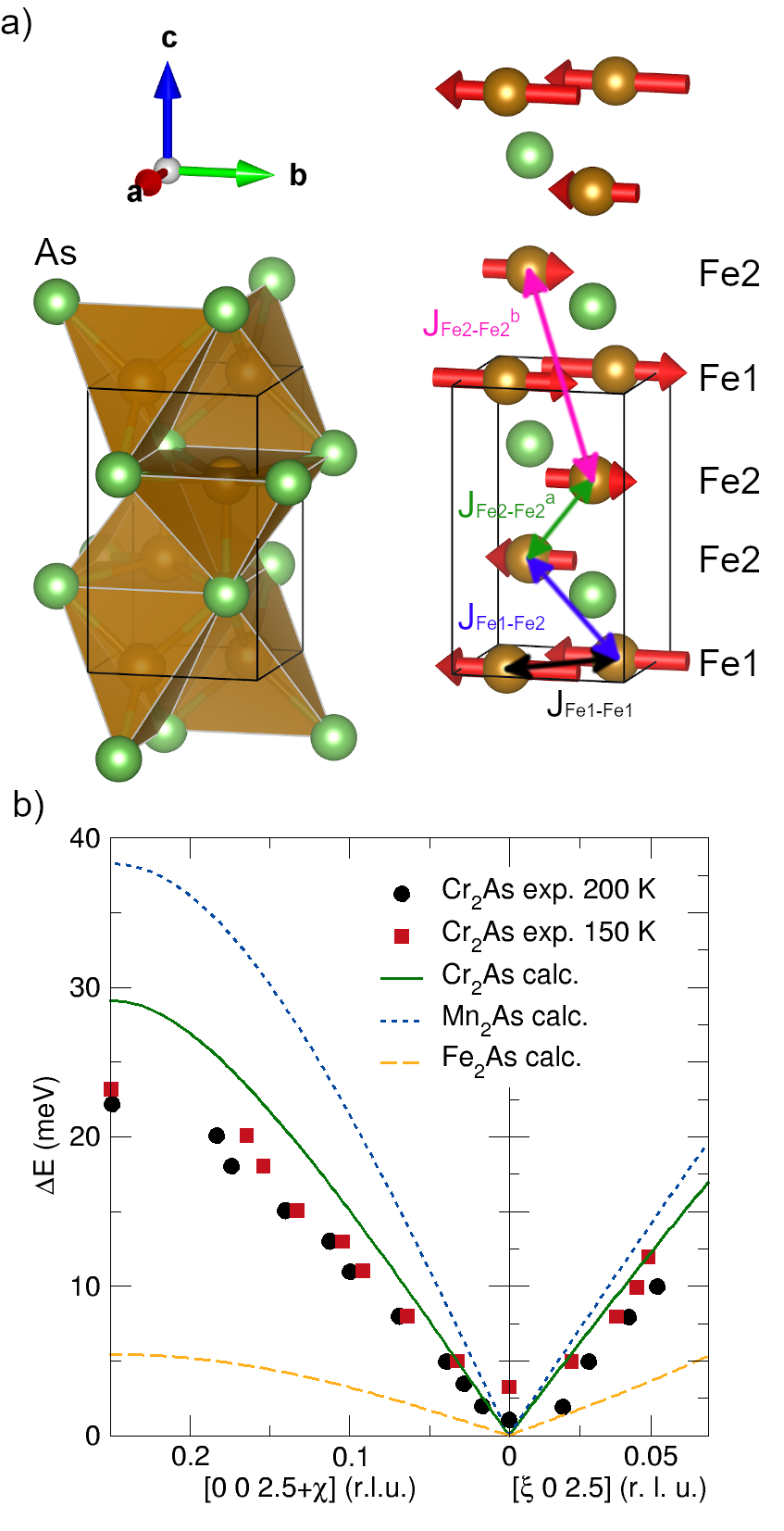} \\
\caption{\label{fig:Cr2As_plot}
The chemical structure of Fe$_2$As (left) showing the FeAs$_4$ tetrahedral and FeAs$_5$ square pyramidal units and the Fe$_2$As magnetic structure (right) with Fe-Fe exchange pathways are shown in (a). Black, blue, green and pink double headed arrows represent Fe1-Fe1, Fe1-Fe2, Fe2-Fe2 nearest-neighbor, and Fe2-Fe2 next-nearest-neighbor interactions, respectively. Comparison between the magnon spectra calculated using the linear spin wave theory from exchange coupling values in reference \citenum{Zhang2013} and the experimental INS values in reference \citenum{Ishimoto1995} are shown in (b) for Cr$_2$As. Also overlaid are the SPRKKR-derived magnon spectra of Mn$_2$As and Fe$_2$As.\cite{Zhang2013} 
} 
\end{figure}

Fe$_2$As contains two different metal atom sites, Fe1 and Fe2, as shown in Figure \ref{fig:Cr2As_plot}(a). Fe1 atoms are centered in FeAs$_4$ tetrahedra, which are arranged to form a square planar grid similar to the anti-PbO type Fe--As layers in iron arsenide superconductors. 
Fe2 atoms form edge-sharing FeAs$_5$ square pyramids. 
Fe$_2$As has a magnetic unit cell that is twice the length of its chemical unit cell along $c$.\cite{Zhang2013,Katsuraki1966}  It is the Fe moments that we are concerned about in the magnon spectrum, but the As contributes to the phonons. The magnetic ground state of Fe$_2$As was determined using single crystal and powder neutron diffraction and consists of alternating slabs of ferromagnetically aligned trilayers of Fe atom planes (Fe2--Fe1--Fe2) as shown in Figure \ref{fig:Cr2As_plot}(a).\cite{Katsuraki1966}
Exchange interactions obtained from  spin polarized, relativistic Korringa-Kohn-Rostoker (SPRKKR)  calculations indicate a strong nearest-neighbor ferromagnetic (FM) Fe1-Fe1 coupling and a weak nearest-neighbor antiferromagnetic (AFM) Fe2-Fe2 interaction.\cite{Zhang2013} 
The Fe-Fe exchange interactions, modeled using SPRKKR calculations, have been explained based on crystal orbital Hamilton population (COHP) curves. The strong Fe1-Fe1 exchange coupling is a result of a strong Fe1-Fe1 anti-bonding orbital overlap as opposed to a weak non-bonding orbital overlap in Fe2-Fe2 nearest neighbor exchange interaction. This case is opposite for Mn$_2$As.\cite{Zhang2013} 
Unlike Fe$_2$As, there is frustration in Mn$_2$As and Cr$_2$As and the magnetic ground state is decided by the dominant exchange interactions.\cite{Zhang2013}


To date, the only direct measurements of exchange interactions in M$_2$As compounds are triple-axis inelastic neutron scattering (INS) measurements on Cr$_2$As single crystals.\cite{Yuzuri1960,Ishimoto1995} 
Magnon spectra calculated from linear spin wave theory using SPRKKR-derived exchange coupling values from Zhang \emph{et al}.\ are plotted on the experimental points from Ishimoto, et al.\ in Figure \ref{fig:Cr2As_plot}(b).\cite{Zhang2013,Ishimoto1995} The experimental magnon spectra roughly agrees with the calculated magnon spectra for the slice plotted in the limited range of reciprocal space. The corresponding magnon spectra for Fe$_2$As and Mn$_2$As from exchange constants in Zhang \emph{et al}.\ are also shown in Figure \ref{fig:Cr2As_plot}(b). 
Since the transition temperature (T$_N$ or T$_C$) is generally proportional to the strength of exchange interactions in a material,\cite{Krishnan2016} the slope of the spin waves along both $H$ and $L$ direction is consistent with T$_N$ of the materials (T$_N$ = 573~K, 393~K and 373~K for Mn$_2$As, Cr$_2$As and Fe$_2$As respectively).\cite{Zhang2013}
Torque magnetometry measurements have been carried out on Fe$_2$As single crystals at different temperatures to determine the four-fold in-plane anisotropy constants.\cite{Yang2020,Achiwa1967} From these measurements, it is clear that the in-plane anisotropy in Fe$_2$As is very small ($<$~1~$\mu$eV) and cannot be resolved using INS measurements. 


Given the technological implications of possible data storage, and the limited momentum space previously examined, a full picture of magnon spectra in metallic antiferromagnets is needed to determine the exchange interactions, and to validate methods of their calculation. Such direct verification has been elusive, and is especially important in highly-correlated 3$d$ systems.
Fe$_2$As single crystals have been grown in centimeter scale,\cite{Katsuraki1966}  making it an ideal candidate to study magnon spectra. In this paper, we report the growth of large Fe$_2$As single crystals and carry out time-of-flight neutron scattering measurements at different temperatures. We identify phonon intensities by comparing with density functional theory-calculated phonon spectra and compare magnon spectra with the reported exchange coupling values. Finally, we refine the exchange coupling values against the INS data to obtain accurate values.

\section{Methods}

\begin{figure}
\centering\includegraphics[width=0.9\columnwidth]{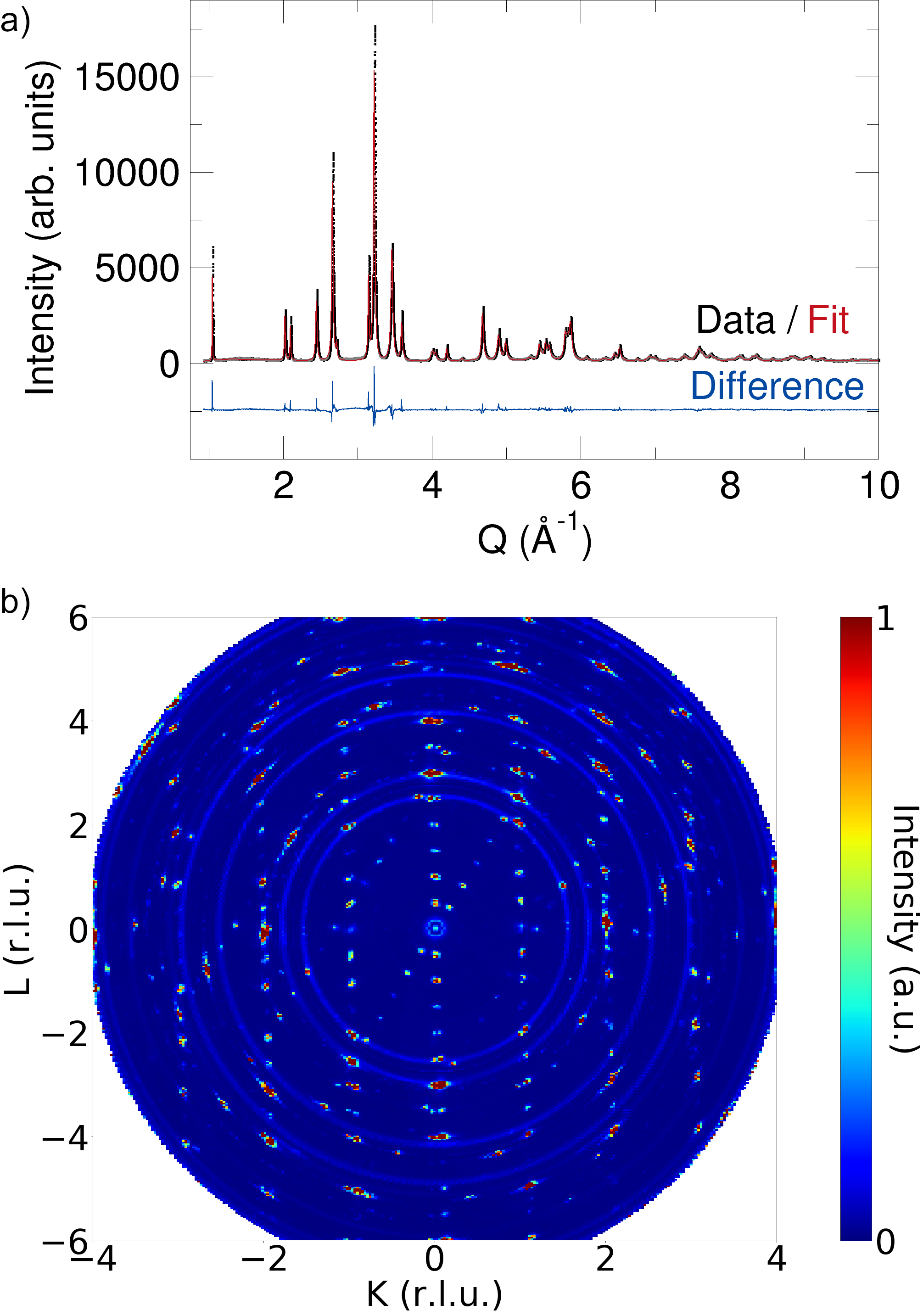} \\
\caption{\label{fig:photo_11_BM}
The Rietveld-refined fit to the synchrotron powder x-ray diffraction data of Fe$_2$As is shown in (a). 
The elastic neutron scattering slice along $K$ and $L$ for $H$ integrated from -0.2 to 0.2 is shown in (b) for $E_i$ = 30~meV. 
}

\end{figure}

\begin{figure*}
\centering\includegraphics[width=0.8\paperwidth]{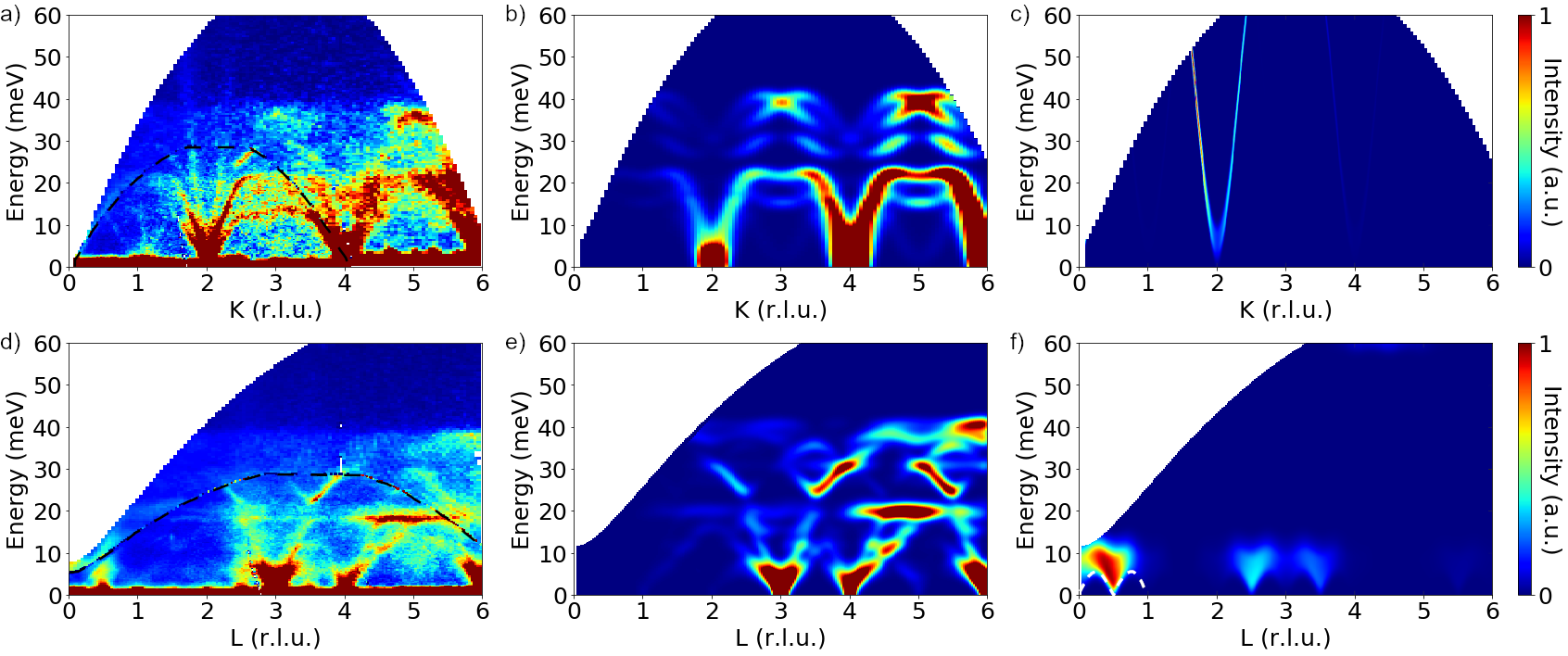} \\
\caption{\label{fig:phonon_magnon_spectra}
INS data of Fe$_2$As measured at 5 K along $K$ with $H$ and $L$ integrated from -0.1 to 0.1 is shown in (a) and along $L$ with $H$ and $K$ integrated from -0.2 to 0.2 is shown in (d). The data with $E_i = 30$~meV (below black dashed lines) have been overlaid on the data with $E_i = 70$~meV in (a) and (d). Panels (b) and (e) show the corresponding simulated phonon spectra obtained from DFT calculations. Panels (c) and (f) show the corresponding simulated magnon spectra derived from exchange constants in reference \citenum{Zhang2013}. The intensities in (b) and (e) have been averaged over 9 equally-spaced phonon spectra in the experimental width along other two $Q$ directions. Similarly, the magnon spectra in (c) and (f) have been averaged over every 0.025 reciprocal lattice units between -0.1 to 0.1 in the other $Q$ directions. The white dashed lines in (f) indicate the calculated magnon spectrum along the [0 0 $L$] direction.
}
\end{figure*}

Large crystals (about 1 cm in length with a mass of about 3 g) were grown from the elements. 
Fe ($>$99.99\% metals basis) and As (99.9999\% metals basis) powders were mixed in 2:1 molar ratio inside an Ar filled glove box and vacuum sealed inside a 7 mm inner diameter quartz tube.
The tube was heated to 600$^{\circ}$C at 1$^{\circ}$C/min and held for 6 hours, heated to 975$^{\circ}$C at 1$^{\circ}$C/min and held for 1 hour, cooled to 900$^{\circ}$C at 1$^{\circ}$C/min and held for 1 hour, then allowed to furnace cool at approximately 10$^{\circ}$C/min to room temperature. The resulting crystals were silver-black in color and produced a mirror like finish when cleaved as shown in Figure S1. The phase purity  was confirmed using synchrotron powder X-ray diffraction  at beamline 11-BM of the Advanced Photon Source in Argonne National Laboratory. Rietveld analysis of the synchrotron data is shown in Figure \ref{fig:photo_11_BM}(a).




The large Fe$_2$As single crystals were gently tapped using a pestle to reveal sharp cleaved surfaces along the $ab$ plane.
Five cleaved crystals of Fe$_2$As, with a total mass of 9 g, were co-aligned onto the base of an Al can and checked with a Multiwire Laue setup at the Spallation Neutron Source (SNS)\cite{mason2006spallation} in Oak Ridge National Laboratory (ORNL). The individual crystals were wrapped in Al foil and sewed to Al shims using Al wires as shown in Figure S2(a) and (b).\cite{supplement} 
One of the five crystals became misaligned, which can be seen in the elastic-scattering slice along $KL$ plane in Figure \ref{fig:photo_11_BM}(b).
Accordingly, regions are selected here from constant energy slices where the effect of the misaligned crystal is minimized. The simulated phonon and magnon spectra do not include the intensity from the misaligned crystal to provide better clarity of the data. Details regarding the intensities from misalignment are provided in Supplementary Materials.\cite{supplement}

The inelastic neutron scattering measurement of Fe$_2$As
was carried out at the ARCS (Wide Angular-Range Chopper Spectrometer) beamline\cite{ doi:10.1063/1.3680104} of the SNS at ORNL. For measurements at  base temperature (about 5~K) and 200~K, the can containing the crystal array was mounted onto a closed cycle refrigerator (CCR) such that the horizontal $(0KL)$ plane was perpendicular to the axis of rotation. For measurement at 400~K (above $T_N$ = 353~K), the crystal array was removed from the can and mounted directly to the CCR. The crystal array was rotated by 360$^{\circ}$ at 1$^{\circ}$ steps in the horizontal plane.
At base temperature, measurements were performed at $E_i = 30$, 70, 200 and 300~meV.  Additional measurements at 70~meV were performed at 200 and 400~K.
Chopper settings were chosen to provide the optimum $Q$ range and resolution conditions, based on Lin, et al. (2019).\cite{Lin2019} For $E_i = 30$ and  70 meV, the 100~meV Fermi chopper was spun at 300 and 480~Hz respectively.  For $E_i = 200$ and 300~meV, the 700~meV chopper was spun at 540 and 420~Hz respectively.  Both choppers have 1.5~mm slit spacing.


Data processing (slicing, folding, and gaussian smoothing) was performed using \textsc{Mantid}.\cite{Arnold2014} The reciprocal lattice units for Fe$_2$As along $K$ (same as $H$) and $L$ correspond to 1.73~\AA$^{-1}$~and 1.05~\AA$^{-1}$, respectively.
Simulated magnon spectra were calculated and refined using the \textsc{SpinW MATLAB} library module, which can solve the spin Hamiltonian using numerical methods and linear spin wave theory.\cite{Toth_2015}
In \textsc{SpinW}, we use a spin-only ($S$)  Hamiltonian based on isotropic exchange interactions $J_{ij}$: $ H = \sum_{i,j} S_i J_{ij} S_j$.
    


Density-functional theory (DFT) calculations were performed using the Vienna \emph{Ab-Initio} Simulation Package \cite{Kresse:1996,Kresse:1999} (VASP). The projector-augmented wave \cite{Blochl:1994} (PAW) scheme was used to describe the electron-ion interaction. Kohn-Sham states are expanded into a plane-wave basis up to a kinetic-energy cutoff of 600 eV. A $15\,\times\,15\,\times\,5$  Monkhorst-Pack (MP) \cite{Monkhorst:1976} $\mathbf{k}$-point grid was used to sample the Brillouin zone. Exchange and correlation was described using the generalized-gradient approximation (GGA) in the formulation by Perdew, Burke, and Ernzerhof.\cite{Perdew:1997} The phonon dispersion was computed with the \textsc{phonopy} package \cite{Phonopy:2015} based on the finite displacement method with total energies from DFT. This calculation used a $3\,\times\,3\,\times\,2$ supercell and a $4\,\times\,4\,\times\,4$ MP $\mathbf{k}$-point grid. The simulated phonon INS spectra were computed using \textsc{OCLIMAX} \cite{Cheng2019} using all phonon eigenvalues from DFT, represented on a reciprocal-space grid. All simulations, in particular all atomic geometry relaxations and phonon dispersion calculations, were performed including noncollinear magnetism and the fully relativistic spin-orbit coupling interaction \cite{Steiner2016}. The instrument parameters used in \textsc{OCLIMAX} correspond to a high resolution measurement at ARCS with an $E_i = 70$~meV.

\begin{figure}
\centering\includegraphics[width=\columnwidth]{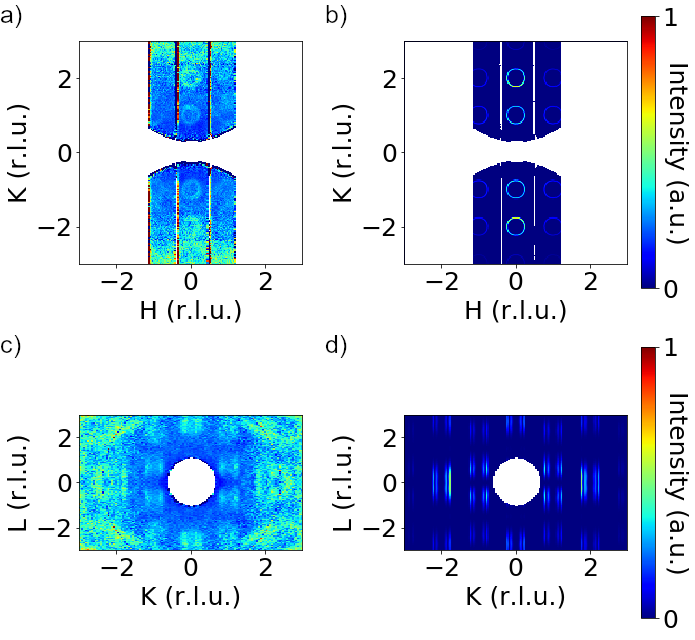} \\
\caption{\label{fig:constant_energy_slices}
Constant-energy INS data reveal magnons most clearly with $E$ integrated from 25 meV to 30 meV for (a) the $H-K$ plane with $L$ integrated from -1 to 1 and (c) $K-L$ plane with $H$ integrated from -0.2 to 0.2 and folded along $L$. Panels (b) and (d) show the corresponding simulated magnon spectra using exchange constants from Zhang \emph{et al}.\cite{Zhang2013} with the same $E$ integration and the orthogonal $Q$ direction summed every 0.1 along the experimental width.
}

\end{figure}

\section{Results and Discussion}

Figures \ref{fig:phonon_magnon_spectra}(a) and \ref{fig:phonon_magnon_spectra}(d) show the inelastic neutron scattering spectra of Fe$_2$As at $T = 5$~K and $E_i =$~70 meV. The corresponding simulated phonon spectra are shown in Figures \ref{fig:phonon_magnon_spectra}(b) and \ref{fig:phonon_magnon_spectra}(e), respectively. Clearly, the phonon contributions form the majority of the experimental spectra, with intensity increasing with $Q$. 
The weak intensity below $E = 10$~meV  at $K = 1$ and $K = 3$ in the experimental data in Figure \ref{fig:phonon_magnon_spectra}(a) is an overlapping phonon band from a misaligned crystal, as seen in Figure \ref{fig:photo_11_BM}(b) and Figure S3(b).\cite{supplement} The group velocities extracted from the three acoustic phonon modes near $\Gamma$ along $K$ (1.215, 2.903, 5.002 km/s) and $L$ (1.745, 1.846, 5.762 km/s) indicate stiffness constants that are the same order of magnitude along  perpendicular directions.

The clearest discrepancy between the experimental spectrum in Figure \ref{fig:phonon_magnon_spectra}(a) and the calculated phonon spectrum in Figure \ref{fig:phonon_magnon_spectra}(b) is the steep excitation arising from $K = 2$. To a first approximation, this magnon mode agrees with the calculated magnon spectrum in Figure \ref{fig:phonon_magnon_spectra}(c), which has a single excitation visible at $K = 2$. When viewed along $a$, the presence of two Fe atoms along $b$ and three Fe atoms along $c$ in the Fe$_2$As chemical unit cell means that the periodicities of the observed phonon and magnon spectra are 2 and 3 along $[0K0]$ and $[00L]$, respectively.

\begin{table*}
\caption{\label{tab:Jvalues} 
Exchange coupling constants (in meV) obtained by fitting the experimental magnon spectra along $K$. 
}
\centering
\begin{tabular}{p{3.5cm}p{2.5cm}p{2.5cm}p{2.5cm}p{2.5cm}p{2.5cm}}
\hline\hline
	&  \bf{Fe1-Fe1 (J$_{Fe1-Fe1}$)} & \bf{Fe1-Fe2 (J$_{Fe1-Fe2}$)} & \bf{Fe2-Fe2 (J$_{Fe2-Fe2^a}$)} & \bf{Fe2-Fe2* (J$_{Fe2-Fe2^b}$)} & \bf{Reduced $\chi^2$} \\

\bf{Distance (\AA)} & 2.547 & 2.6859 & 3.2774 & 4.7160\\
\hline\hline
\bf{Zhang \emph{et al}.} & -25.4 & -6.52 & 3.52 & -8.52 & 54.55\\
\hline
\bf{Fit} & -48.37(25) & -4.42(25) & 5.16(12) & -8.52 & 6.47\\
\hline\hline
\end{tabular}
~\\
\end{table*}

From DFT SPRKKR-derived exchange coupling values in Zhang \emph{et al}.,\cite{Zhang2013} magnon spectra were calculated using the linear spin wave theory and simulated with an energy binning of 3~meV, which corresponds to our experimental resolution near the elastic limit with $E_i$~=~70~meV. Figures \ref{fig:phonon_magnon_spectra}(c) and \ref{fig:phonon_magnon_spectra}(f) show the magnon spectra along $K$ and $L$ directions, respectively. All the intensities in Figures \ref{fig:phonon_magnon_spectra}(a) and \ref{fig:phonon_magnon_spectra}(d) are accounted for in the simulated phonon and magnon spectra. The spectral weight of the magnons is mostly negligible along $L$ except for the locations shown in Figure \ref{fig:phonon_magnon_spectra}(f). Constant-energy slices at $E$~=~25~meV in the $H-K$ and $K-L$ planes are shown in Figure \ref{fig:constant_energy_slices}(a,c). The simulated magnon spectra in Figure \ref{fig:constant_energy_slices}(b,d) give excellent reproduction of the corresponding INS data. Smaller magnon circles in Figure \ref{fig:constant_energy_slices}(a) as compared to the ones in Figure \ref{fig:constant_energy_slices}(b) indicate the possibility of stronger in-plane exchange interactions than those reported in Zhang \emph{et al}.\cite{Zhang2013}

On quick inspection of Figure \ref{fig:phonon_magnon_spectra}(c), the energy dependence along $K$ appears to be a simple 1-D Heisenberg FM spin chain where the magnon spectrum varies as $1 - \textrm{cos} (Ks)$,\cite{Stancil} $s$ being the interatomic spacing for the FM chain along $b$. Since the spins in Fe$_2$As are all aligned parallel to each other along $b$, the exchange interactions are consistent with the ground state. However, the spectrum is repeated every two reciprocal lattice units along $K$ since the unit cell contains two Fe atoms along $b$. The magnon spectrum along $L$ in Figure \ref{fig:phonon_magnon_spectra}(f) has a similar $|\textrm{sin}(Ls)|$ dependence as seen in a 1-D Heisenberg AFM spin chain where $s$ is the interatomic spacing for the AFM chain along $c$. Unlike a 1-D Heisenberg AFM spin chain, however, Fe$_2$As contains AFM-stacked trilayers of Fe atoms. 
The dispersion of the spin waves in Figure \ref{fig:phonon_magnon_spectra}(a,d) indicate a strong FM coupling along $b$ and weak trilayer AFM coupling along $c$ as also confirmed from the exchange coupling values in Zhang et. al. (2013)\cite{Zhang2013} in Table \ref{tab:Jvalues}.

From torque magnetometry measurements in the $ab$ plane, the four-fold in-plane anisotropy in Fe$_2$As at liquid nitrogen temperatures was reported to be around 700~erg/g, which is~0.3 $\mu$eV/cell.\cite{Achiwa1967}
Recent measurements at 5~K conclude that this quantity is much lower than previously reported at 0.074~$\mu$eV/cell (150 J/m$^3$) and it deceases to zero at around 150 K.\cite{Yang2020}
The out-of-plane 2-fold anisotropy value was estimated using DFT calculations to be 410 $\mu$eV/cell (-830 kJ/m$^3$).\cite{Yang2020} A similarly small anisotropy was reported for CuMnAs using relativistic calculations where the in-plane anisotropy was calculated to be less than 1 $\mu$eV/cell and the out-of-plane value was reported to be 127 $\mu$eV/cell.\cite{Wadley2015} Our ARCS experimental resolution in $E$ near the elastic limit is around  3 - 5\% of $E_i$, so anisotropy in Fe$_2$As can be neglected. 


The calculated magnon spectra using exchange constants from Zhang \emph{et al}.\cite{Zhang2013} underestimate the magnon energy along $K$ (by about 24\% at $K = 1.25$).
Ideally, refinement of the magnon spectra with \textsc{SpinW}\cite{Toth_2015} should extract more accurate exchange constant values.
Along $L$, as shown in Figure \ref{fig:phonon_magnon_spectra}(f), even small integration of $Q$ in the orthogonal directions causes significant bleeding over of intensity due to the steep magnon modes in the $H$ and $K$ directions. The same effect is seen for $K = 1$, shown in Figure S4(c).\cite{supplement} 
Hence, the calculated magnon spectra in Figure \ref{fig:phonon_magnon_spectra}(f) was assumed to be correct and points were taken from the calculated magnon spectra along $L$. This ensures a net weak AFM coupling along $L$ for the purpose of refinement.
Higher-energy INS data collected at 5~K using $E_i=200$~meV and 300~meV are shown in Figures S5(a,b).\cite{supplement} As shown in Figures S5, we see that the scattering extends up beyond 120~meV. We did not use this data in the fits as the itinerant nature of the moments at this energy leads to significant damping that blurs the mode position.  Nevertheless the results obtained from the fits are consistent with this scattering. Only the INS data obtained from $E_i=30$~meV and 70~meV  were considered for refinement.
From high temperature susceptibility measurements of Fe$_2$As,\cite{Katsuraki1966} the effective total moment per Fe is estimated as 4.66~$\upmu_\text{B}$ averaged over the two sublattices. The ordered moment, which is estimated by neutron diffraction in Fe1~=~0.95~$\upmu_\text{B}$ and Fe2~=~1.52~$\upmu_\text{B}$, is lower than 4.66~$\upmu_\text{B}$.\cite{Katsuraki1966,Zhang2013} So, the rest of the moment can be assumed to be itinerant or short-ranged. The extracted average total moment of the Fe sublattices seems unusually high and well-calibrated high temperature susceptibility measurements are thus warranted.
The set of experimental data points used to refine the exchange interactions is shown in Figure S6.\cite{supplement} Data points were collected by making horizontal line cuts across the magnon spectra along $K$. Vertical line cuts were dominated by the flatter phonon modes. Hence, the standard deviation of energy for the purpose of refinement was assumed to be a constant of 1~meV.



\begin{figure}
\centering\includegraphics[width=\columnwidth]{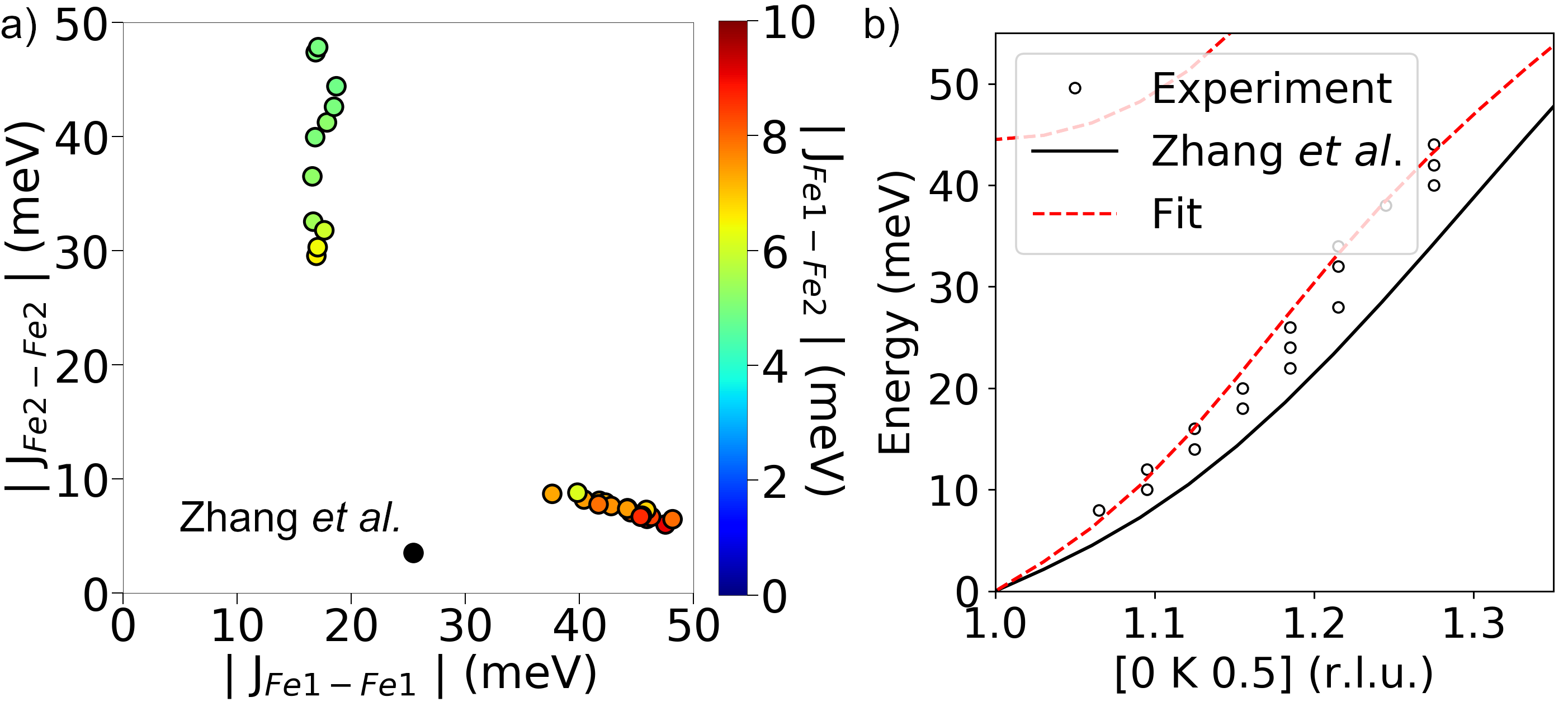} \\
\caption{\label{fig:zhang_spinw_refinement}
The result of unconstrained optimization of the exchange coupling values when only three nearest-neighbor interactions are considered is shown in (a). The reduced $\chi^2$ values of all points are less than 7, but these three-$J_{ij}$ fits are disallowed by intensity mismatches to the INS data. In (b), comparison of the fit of a four-$J_{ij}$ model obtained by fixing the NNN Fe2-Fe2 interaction to be $-8.52$~meV and the calculated magnon spectra from the exchange constants from Zhang \emph{et al}.\cite{Zhang2013} leads to an improvement of the fit, with much larger Fe1-Fe1 interaction (see Table \ref{tab:Jvalues}). 
}
\end{figure}

Fe$_2$As is expected to contain a strong Fe1-Fe1 exchange interaction due to a strong anti-bonding interaction as seen in crystal orbital Hamilton population curves.\cite{Zhang2013}
The Fermi level crosses a narrow band along the $X$--$R$ Brillouin zone boundary. Weak Fe2-Fe2 interaction is expected due to the weak antibonding $xy$ and $xz$ orbital overlap at point $R$. However, there is a significant overlap of the Fe2 and As orbitals indicating a possibility of strong superexchange interaction.\cite{Zhang2013} The Fe1-Fe1 and Fe1-Fe2 nearest neighbor exchange interactions can be attributed to direct exchange and the nearest-neighbor and next-nearest-neighbor (NNN) Fe2-Fe2 exchange interactions can be attributed to indirect exchange although there is some direct exchange also possible in the nearest neighbor Fe2-Fe2 exchange interactions.\cite{Zhang2013}  Strong indirect exchange interactions have been reported for MnFeAs, another compound in the Cu$_2$Sb structure type, using SPRKKR calculations.\cite{Zhang2015} From the study of MnFeAs, we can say that there are two possible contributions to the indirect exchange interactions in this material. One effect is due to superexchange interactions mediated by As atoms and the other effect arises from RKKY interactions due to the compound being metallic.\cite{Zhang2015}


The smallest number of exchange coupling constants required to produce magnon modes along $L$ are the Fe1-Fe2 and Fe2-Fe2 nearest-neighbor interactions. However, the fit is poor (reduced $\chi^2$ = 9.03) and is greatly improved upon adding a third $J_{ij}$, the other nearest-neighbor exchange interaction Fe1-Fe1. 
The refinement with three $J_{ij}$ was carried out using the particle swarm optimization technique with a limit of 20 iterations.
Selecting points having reduced $\chi^2 < 7$  from the result of 50 runs, Figure \ref{fig:zhang_spinw_refinement}(a) shows the exchange constants obtained when the magnon spectra is refined to a model containing only the three nearest-neighbor exchange interactions. 
We can roughly divide the points into two clusters. The cluster of exchange coupling values with strong Fe2-Fe2 nearest-neighbor interactions are incorrect since we know from previous computational studies that Fe$_2$As should have nearest neighbor strong Fe1-Fe1 coupling and a weak Fe2-Fe2 coupling.\cite{Zhang2013} Also, the intensity of the magnon modes in the simulated magnon spectra for this set of $J_{ij}$ arising from [0 1 0.5] is weak, as shown in Figure S7(a),\cite{supplement} which is invalidated by the experimental data. In the other cluster, the Fe1-Fe1 nearest neighbor exchange coupling seems much higher than the reported value of 25.4~meV. However, the simulated magnon spectra from any point in that three-$J_{ij}$ cluster shows that the magnon spectra becomes mostly flat above 60~meV and also drops down below 60~meV near $K = 1$ and 2 as shown in Figure S7(b).\cite{supplement} This is not seen in the experimental magnon spectra. The addition of a fourth $J_{ij}$ is necessary to prevent the magnon spectra from flattening at high energies. Similar to Zhang \emph{et al}.,\cite{Zhang2013} we can choose the NNN Fe2-Fe2 exchange interaction as the fourth exchange interaction for refinement.

The effect of adding a NNN Fe2-Fe2 exchange interaction is mainly at higher energies where the experimental spectra are unresolved. Thus a fourth $J_{ij}$ is necessary, but not refinable from INS data. We fixed the value of the Fe2-Fe2 NNN exchange interaction to that of Zhang \emph{et al}.\cite{Zhang2013} and the remaining three nearest-neighbor exchange interactions were refined 50 times. Four of the runs converged to a reduced $\chi^2 \approx 6.5$, as compared to $\chi^2 > 9$ for the rest of the runs. The mean exchange coupling value from the four runs is shown in Table \ref{tab:Jvalues} and the calculated magnon spectrum using linear spin wave theory is plotted in Figure \ref{fig:zhang_spinw_refinement}(b). We can see that the Fe1-Fe1 nearest-neighbor exchange interaction is much stronger than the SPRKKR value, which was also seen in the earlier model with only three nearest-neighbor exchange interactions. One should note that, for the sake of optimization, an upper limit of 50~meV was kept for all exchange coupling constants. The value for Fe1-Fe1 exchange coupling is close to this limit. Given that the Fe1-As bond is shorter than one of the Fe2-As bonds, it is possible that there is also some superexchange component in the NNN Fe1-Fe2 interaction. The Fe1-Fe2 distance of 4.4~\AA\ is also shorter than the NNN Fe2-Fe2 distance (4.4716~\AA), allowing for possible RKKY interactions. Although we do not have enough experimental data to elucidate the role of this exchange interaction, it may not be neglected.

If AF materials are to be used in future MRAM devices, it is essential that the 4-fold in-plane anisotropy values surpass 10~meV so that the domains are stable at operating temperatures. Unlike CuMnAs, Fe$_2$As is complicated by the presence of two different magnetic atom sites with different point groups. When the current is parallel to the N\'eel vector, the effective fields on the two Fe sublattices from the field-like torque are perpendicular to each other and the strength of the Fe1-Fe2 exchange interaction may play a role in the electrical switching of the N\'{e}el vector. Hence, it is important that we are able to predict and measure these interactions accurately. Similar to refining the magnon spectra from the experiment, the exchange coupling values obtained from SPRKKR calculations are also contingent on the chosen model. Exchange interactions obtained from ab-initio calculations are known to give largely different values than the experiment, as seen in the case of Mn$_3$Sn.\cite{Park2018} Hence, a more robust determination of exchange energies is warranted. Future efforts could be aided by developing the capability to refine these values while considering magnon intensity quantitatively, and by evaluating metallic antiferromagnets where the higher-energy magnon dispersion is experimentally resolvable. 


\section{Conclusions}

The experimental phonon spectra of Fe$_2$As matches the simulated phonon spectra from DFT calculations very well. The simulated magnon spectra calculated using exchange coupling values from Zhang \emph{et al}. agrees qualitatively with the experimental magnon spectra. The energy values are underestimated by about 20\% along $K$ direction. The anisotropy values were deemed small enough to be neglected for the purpose of refinement and the magnon spectra was refined using a Heisenberg Hamiltonian. For the model used in Zhang \emph{et al}., keeping the value of Fe2-Fe2 nearest neighbor interaction to be a constant, the Fe1-Fe1 nearest neighbor exchange interaction was estimated to be much stronger than previously calculated. 
The in-plane and out-of-plane phonon group velocities are the same order of magnitude, but the magnetic interactions are strongly 2D in nature. This shows that the 2D nature of the magnetism does not arise from weak out-of-plane bonding.


\section{Acknowledgments}

This work was undertaken as part of the Illinois Materials Research Science and Engineering Center, supported by the National Science Foundation MRSEC program under NSF Award No. DMR-1720633. The characterization was carried out in part in the Materials Research Laboratory Central Research Facilities, University of Illinois. This work made use of the Illinois Campus Cluster, a computing resource that is operated by the Illinois Campus Cluster Program (ICCP) in conjunction with the National Center for Supercomputing Applications (NCSA) and which is supported by funds from the University of Illinois at Urbana-Champaign. This research is part of the Blue Waters sustained-petascale computing project, which is supported by the National Science Foundation (Awards No. OCI-0725070 and No. ACI-1238993) and the state of Illinois. Blue Waters is a joint effort of the University of Illinois at Urbana-Champaign and its National Center for Supercomputing Applications. This research used resources of the Spallation Neutron Source, a DOE Office of Science User Facility operated by Oak Ridge National Laboratory, and the Advanced Photon Source, a DOE Office of Science User Facility operated for the DOE Office of Science by Argonne National Laboratory under Contract No. DE-AC02-06CH11357. The authors thank Yan Wu, Huibo Cao and Douglas Abernathy for helpful discussions regarding the experiment.

\bibliography{INS_Fe2As}

\end{document}



\begin{center}
\Large 
\textbf{Strongly two-dimensional exchange interactions in the in-plane metallic antiferromagnet Fe$_2$As probed by inelastic neutron scattering}\\
\vspace{1em}
Supplementary Material\\
\vspace{1em}
\normalsize
Manohar H. Karigerasi, Kisung Kang, Garrett E. Granroth, Arnab Banerjee, Andr\'e Schleife, Daniel P. Shoemaker
\end{center}

Information on T0 chopper used for measurements

The T0 chopper blocks the prompt pulse and additional openings of the Fermi chopper.  For E$_i = 30$, 70 and 200~meV, 90~Hz was used. For E$_i = 300$~meV, 120~Hz was used.
For the 400~K 70~meV run, the T0 chopper was at 60~Hz and provided equivalent performance.

\vspace{2em}

\begin{figure}[h]
\centering\includegraphics[width=\columnwidth]{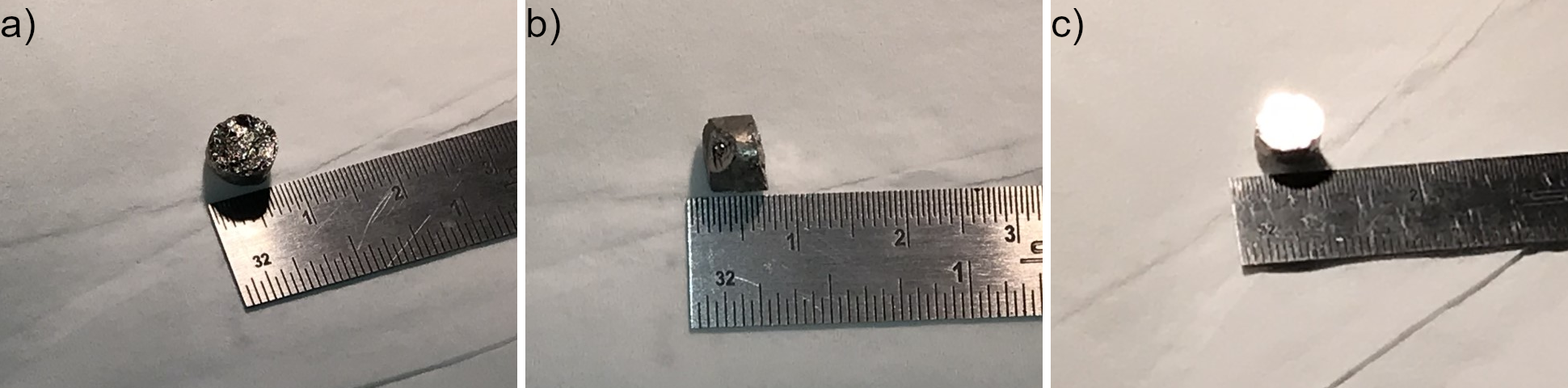} \\
\caption{\label{fig:crystal_array}
One of the Fe$_2$As crystal used for the inelastic neutron scattering measurement with the cleaved surface facing up is shown in (a) and facing to the right side is shown in (b). (c) shows the mirror-like metallic lustre of Fe$_2$As crystals in the cleaved surface when seen from an angle. The values displayed on the scale closest to the sample are in centimeters.
} 
\end{figure}

\begin{figure}[h]
\centering\includegraphics[width=0.5\columnwidth]{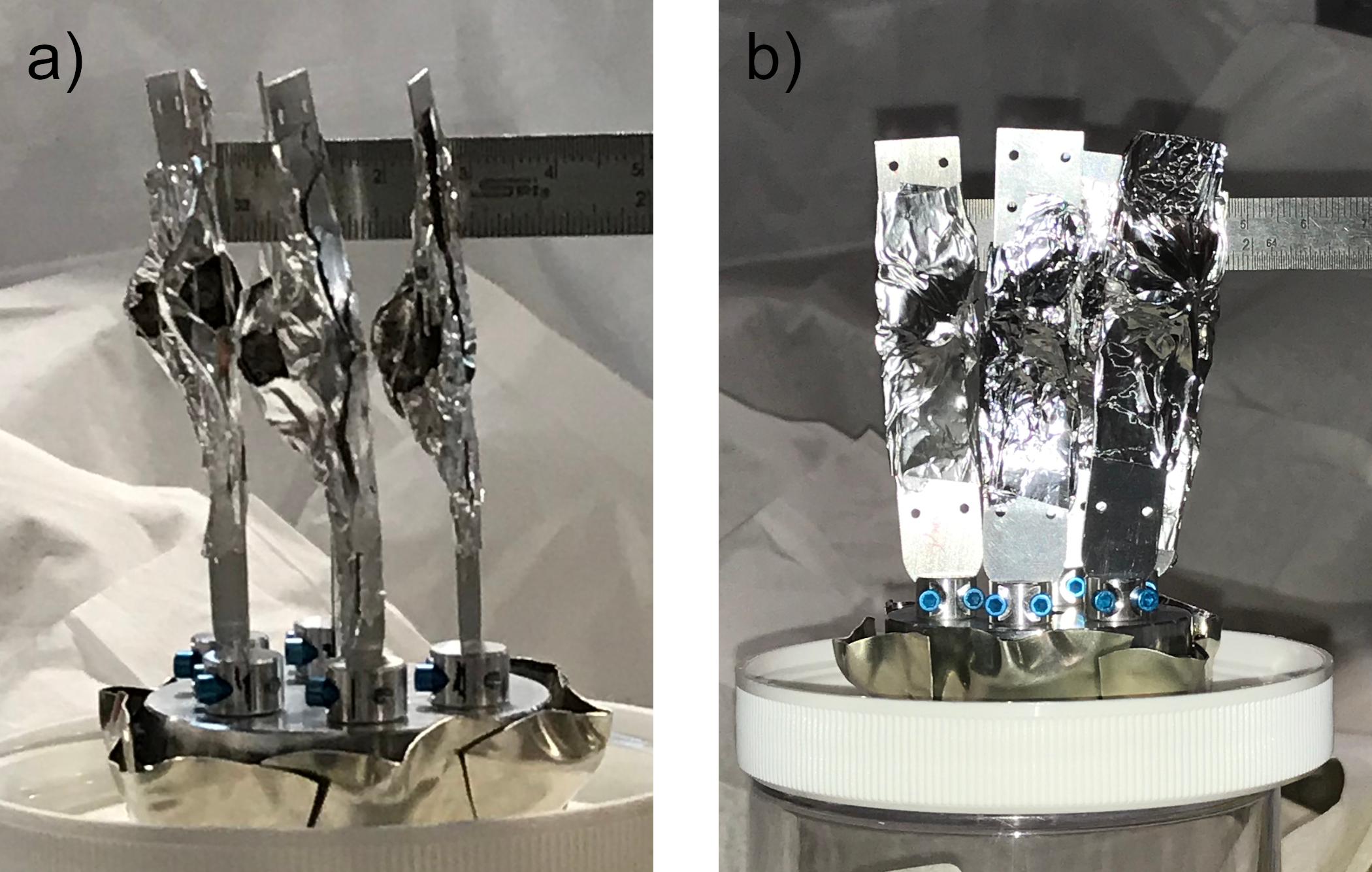} \\
\caption{\label{fig:crystal_array}
Five Fe$_2$As single crystals (about 9 g), wrapped in Al foil and co-aligned using the Laue instrument is shown facing (a) $c$ axis of crystal and facing (b) $b$ axis of crystal.
} 
\end{figure}

\begin{figure}[h]
\centering\includegraphics[width=0.8\columnwidth]{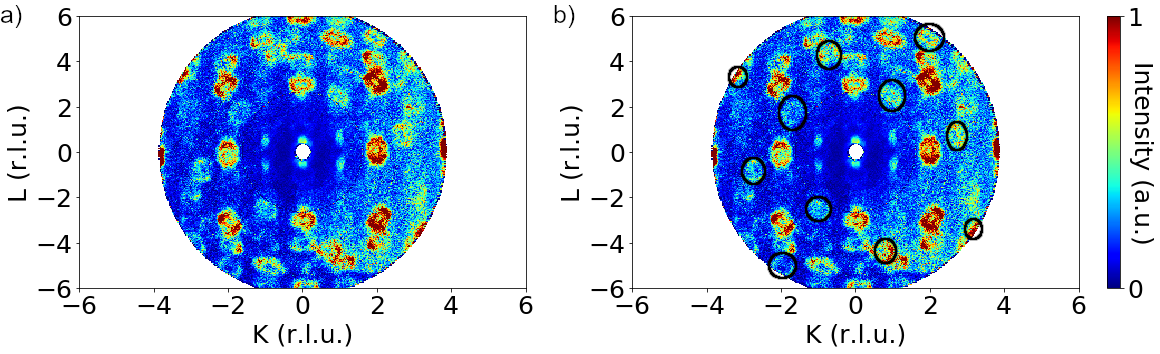} \\
\caption{\label{fig:misaligned_elastic}
A constant energy slice along the $K-L$ plane with energy integrated from 6~meV to 8~meV using $E_i$~=~30 meV is shown in (a). Visible intensity from the misaligned crystal is circled in (b). The misaligned crystal has significant effect on phonons at higher $Q$.
}
\end{figure}

\begin{figure}[h]
\centering\includegraphics[width=\columnwidth]{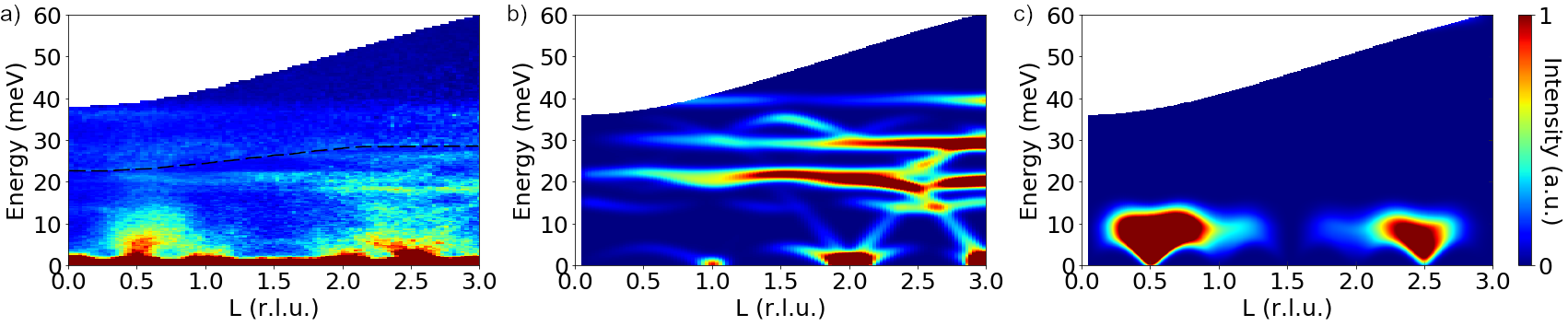} \\
\caption{\label{fig:phonon_magnon_spectra_01L}
INS data along $L$ with $H$ integrated from -0.1 to 0.1 and $K$ from -0.9 to 1.1 is shown in (a). Data from E$_i$~=~30~meV (below dashed lines) has been overlaid on top of data from E$_i$~=~70~meV. The corresponding simulated phonon spectra along [0 1 $L$] is shown in (b). (c) shows the magnon spectra along $L$ where $H$ and $K$ have been averaged every 0.025 units in the experimental width.
}
\end{figure}

\begin{figure}[h]
\centering\includegraphics[width=\columnwidth]{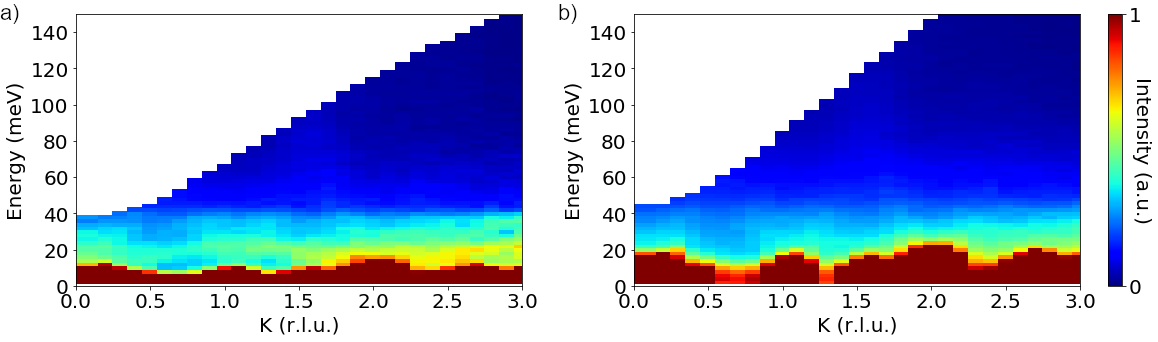} \\
\caption{\label{fig:high_energy_data}
The magnon spectra along $K$ is shown using $E_i$ = (a) 200~meV and (b) 300~meV where $H$ is integrated from -0.2 to 0.2 and $L$ is integrated from -1 to 1.
}
\end{figure}

\begin{figure}[h]
\centering\includegraphics[width=0.6\columnwidth]{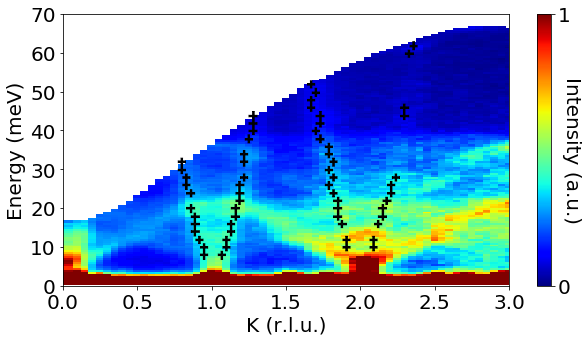} \\
\caption{\label{fig:exp_data_points_01L}
The experimental data points used for refinement have been overlaid on top of the corresponding magnon spectra along $K$ obtained using $E_i$~=~70~meV. $H$ is integrated from -0.2 to 0.2 and $L$ is integrated from 0.3 to 0.7.
} 
\end{figure}

\begin{figure}[h]
\centering\includegraphics[width=\columnwidth]{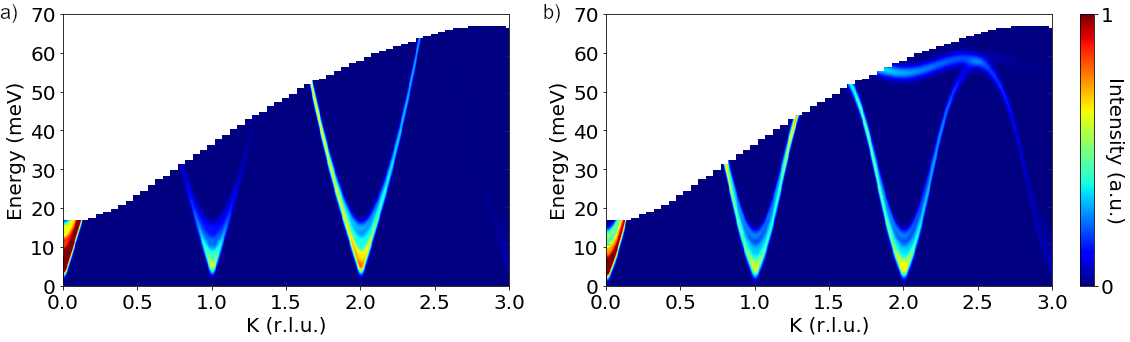} \\
\caption{\label{fig:3J_spinw_refinement}
For a model containing only the three nearest neighbor interactions, the simulated magnon spectra calculated along $K$ with $H$ and $L$ averaged every 0.025 units between 0.0 to 0.1 and 0.5 to 0.6 respectively for (a) a point in the cluster having high Fe2-Fe2 nearest neighbor exchange interaction and low Fe1-Fe1 nearest neighbor exchange interaction and (b) a point in the cluster with small Fe2-Fe2 nearest neighbor exchange interaction but large Fe1-Fe1 nearest neighbor exchange interaction.
}
\end{figure}